\documentclass[a4paper,conference]{IEEEtran}
\IEEEoverridecommandlockouts
\usepackage{cite}
\usepackage{amsmath,amssymb,amsfonts}
\usepackage{algorithm}
\usepackage{algorithmic}
\usepackage{graphicx}
\usepackage{textcomp}
\usepackage{xcolor}

\usepackage[acronym]{glossaries}
\usepackage{scalerel}
\usepackage{tikz}
\usepackage{svg}
\usepackage{hyperref} 
\usepackage{booktabs}
\usepackage{multirow}
\usepackage{siunitx}
\usepackage{makecell} 
\usepackage{array}    

\def\showauthors{1} 
\def\preprint{0} 

\if\preprint1
\usepackage[all]{background}
\usepackage{stackengine}
\fi

\begin{document}

\if\preprint1
\setstackEOL{\\}
\setstackgap{L}{\normalbaselineskip}
\SetBgContents{\color{gray}{\tiny \Longstack{PREPRINT - Accepted at the 2025 IEEE Computer Society Annual Symposium on VLSI (ISVLSI 2025)}}}
\SetBgPosition{4.5cm,1cm}
\SetBgOpacity{1.0}
\SetBgAngle{0}
\SetBgScale{1.8}
\fi

\title{Evaluating the Scalability of Binary and Ternary CNN Workloads on RRAM-based Compute-in-Memory Accelerators}

\if\showauthors1
\author{
  \IEEEauthorblockN{
    José Cubero-Cascante,
    Rebecca Pelke,
    Noah Flohr,
    Arunkumar Vaidyanathan,\\
    Rainer Leupers,
    Jan Moritz Joseph
  }
  \IEEEauthorblockA{
    \textit{Institute for Communication Technologies and Embedded Systems, RWTH Aachen University}, Aachen, Germany \\
    \{cubero,pelke,flohr,vaidyanathan,leupers,joseph\}@ice.rwth-aachen.de
  }
  \thanks{
    This work was funded by Germany's Federal Ministry of Education and Research (BMBF) in the project NEUROTEC II (Project Nos. 16ME0398K, 16ME0399).
    Copyright 979-8-3315-3477-6/25/\$31.00 ©2025 IEEE.
    }
}
\else
\author{
  \IEEEauthorblockN{Authors are removed for submission version}
  \\
  \\
  \IEEEauthorblockA{Affiliations are removed for submission version}
  \\
  \\
}
\fi

\newacronym{esl}{ESL}{Electronic System Level}
\newacronym[plural=ICs,firstplural=Integrated Circuts (ICs)]{ic}{IC}{Integrated Circuit}
\newacronym[plural=KPIs,firstplural=Key Performance Indicators (KPIs)]{kpi}{KPI}{Key Performance Indicator}
\newacronym[plural=DNNs,firstplural=Deep Neural Networks (DNNs)]{dnn}{DNN}{Deep Neural Network}
\newacronym[plural=CNNs,firstplural=Convolutional Neural Networks (CNNs)]{cnn}{CNN}{Convolutional Neural Network}
\newacronym[plural=QNNs,firstplural=Quantised Neural Networks (QNNs)]{qnn}{QNN}{Quantised Neural Network}
\newacronym[plural=BNNs,firstplural=Binary Neural Networks (BNNs)]{bnn}{BNN}{Binary Neural Network}
\newacronym[plural=TNNs,firstplural=Ternary Neural Networks (TNNs)]{tnn}{TNN}{Ternary Neural Network}
\newacronym[plural=MVMs,firstplural=Matrix Vector Multiplications (MVMs)]{mvm}{MVM}{Matrix Vector Multiplication}
\newacronym{rram}{RRAM}{Resistive Random Access Memory}
\newacronym{reram}{ReRAM}{Redox-based Random Access Memory}
\newacronym{pcm}{PCM}{Phase Change Memory}
\newacronym[plural=ADCs,firstplural=Analog-to-Digital Converters (ADCs)]{adc}{ADC}{Analog-to-Digital Converter}
\newacronym[plural=DACs,firstplural=Digital-to-Analog Converters (DACs)]{dac}{DAC}{Digital-to-Analog Converter}
\newacronym{soc}{SoC}{System-on-Chip}
\newacronym{cim}{CIM}{Compute-in-Memory}
\newacronym{mac}{MAC}{Multiply-Accumulate}
\newacronym{kcl}{KCL}{Kirchhoff's Circuit Laws}
\newacronym{itrs0}{ITRS}{International Technology Roadmap for Semiconductors}
\newacronym{lrs}{LRS}{Low Resistive State}
\newacronym{hrs}{HRS}{High Resistive State}

\maketitle

\begin{abstract}
The increasing computational demand of Convolutional Neural Networks (CNNs) necessitates energy-efficient acceleration strategies.
Compute-in-Memory (CIM) architectures based on Resistive Random Access Memory (RRAM) offer a promising solution by reducing data movement and enabling low-power in-situ computations.
However, their efficiency is limited by the high cost of peripheral circuits, particularly Analog-to-Digital Converters (ADCs).
Large crossbars and low ADC resolutions are often used to mitigate this, potentially compromising accuracy.
This work introduces novel simulation methods to model the impact of resistive wire parasitics and limited ADC resolution on RRAM crossbars.
Our parasitics model employs a vectorised algorithm to compute crossbar output currents with errors below 0.15\% compared to SPICE.
Additionally, we propose a variable step-size ADC and a calibration methodology that significantly reduces ADC resolution requirements.
These accuracy models are integrated with a statistics-based energy model.
Using our framework, we conduct a comparative analysis of binary and ternary CNNs.
Experimental results demonstrate that the ternary CNNs exhibit greater resilience to wire parasitics and lower ADC resolution but suffer a 40\% reduction in energy efficiency.
These findings provide valuable insights for optimising RRAM-based CIM accelerators for energy-efficient deep learning.
\end{abstract}


\section{Introduction}

\glspl{cnn} are a cornerstone of modern artificial intelligence, excelling in tasks like image recognition, object detection, and video analysis~\cite{lecun_deepl_2015}.
However, these applications demand very high computational power,
with a significant portion of the energy consumed in transferring learned weights between non-volatile storage and processing units like CPUs or Deep Learning Accelerators (DLAs)~\cite{andrulis_cimloop_2024,wei_lu_CIM_DNN_review_2024}.

\gls{cim} has emerged as a promising means for minimising those costly data movements.
Moreover, by leveraging in-situ analog computations with emerging non-volatile memristive technologies, low-energy \glspl{mvm} can be achieved~\cite{sun_nature_full_spectrum_2023}.
Various memristive devices are being explored for this purpose, including:
\gls{reram}~\cite{yao_fully_hw_memr_cnn,prezioso_training_2014},
\gls{pcm}~\cite{joshi_IBM_accurate_pcm_2020},
charge-trapping~\cite{chen_2024_perovskite} and floating-gate~\cite{fuller_ifg}.

\begin{figure}[!t]
  \begin{centering}
      \includegraphics[width=0.84\columnwidth]{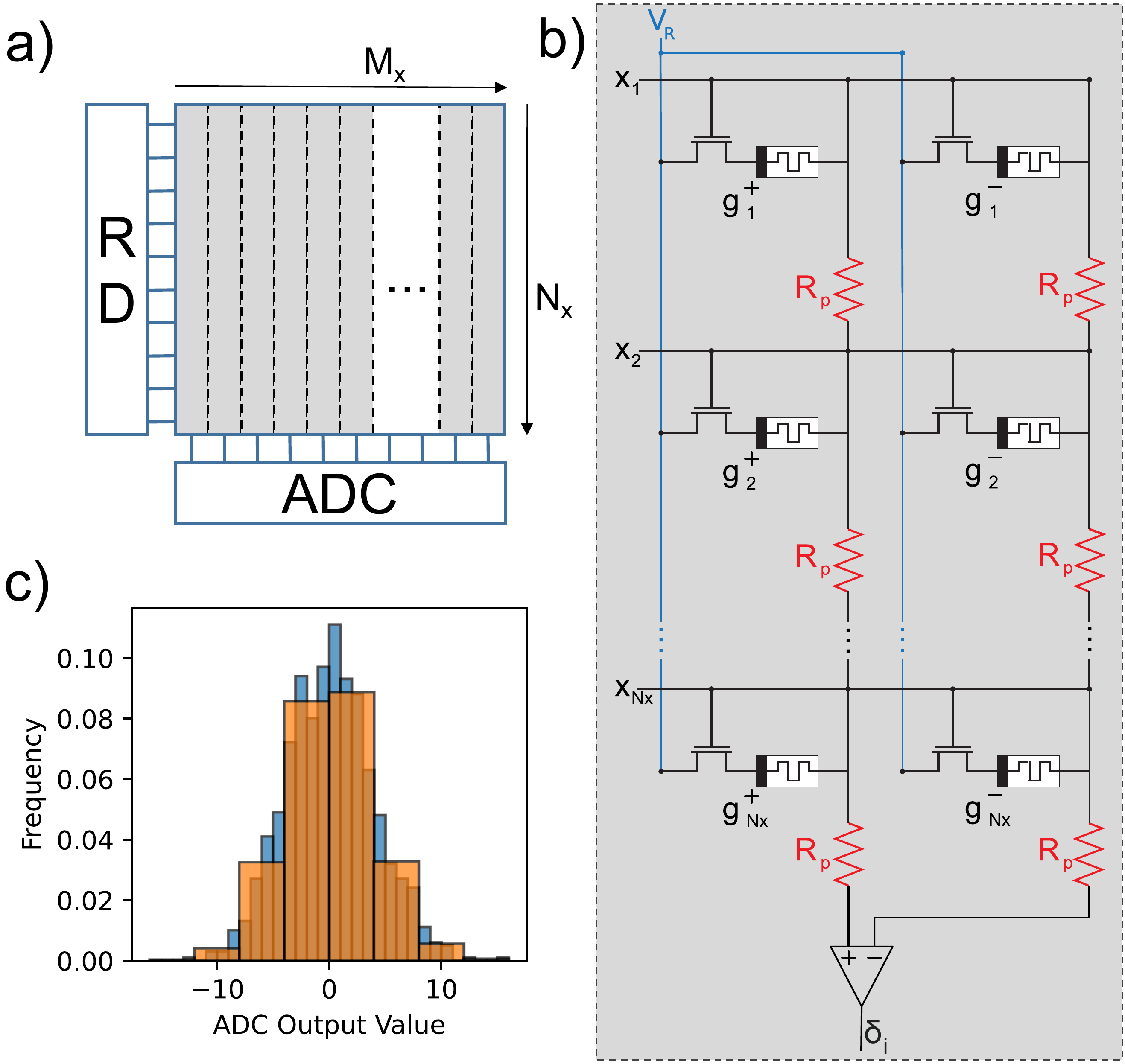}
      \caption{
        Overview of the target RRAM-based CIM accelerator and the scalability challenges.
        a) CIM-core with a $N_X \times M_X$ RRAM crossbar, Row Drivers (RD) and ADC.
        b) Differential column dot-product circuit with wire parasitic resistances $R_p$ highlighted in red.
        c) ADC output histograms showing the accuracy loss caused by a coarse quantisation step $\Delta_Q$.
      }
      \label{fig:crossbar}
  \end{centering}
\end{figure}

However, analog CIM solutions face a significant bottleneck in the form of power-hungry \glspl{adc}, which severely constrain their overall energy efficiency.
To amortise the cost of these peripherals, two competing strategies are commonly employed: extending the size of the crossbar arrays and reducing the ADC resolution~\cite{andrulis_cimloop_2024}.
These strategies come at the expense of accuracy loss, as larger crossbars exacerbate signal degradation due to wire parasitic resistance~\cite{cao_parasitic-aware_2022,xiao_sandia_parasitics_2021}.
Furthermore, reduced ADC resolution leads to either clipping or coarser quantisation.
Hence, understanding the scalability of \gls{rram}\footnote{We use RRAM as an umbrella term covering all memristive technologies.} crossbars remains a critical open research question.

This work addresses the scalability question with new simulation models that quantify the effect of crossbar size, ADC resolution and wire parasitics on accuracy and energy efficiency.
We focus specifically on binary 1T1R crossbars and applications based on \glspl{bnn} and \glspl{tnn},
which are particularly well-suited for low-power devices.

Our main contributions are:
\begin{itemize}
\item a novel model for quantifying the effect of wire parasitics on RRAM crossbars,
\item a simple yet efficient calibration methodology for reducing ADC resolution,
\item an integrated statistics-based energy model,
\item a comprehensive comparison of the impact of wire parasitics and ADC resolution on BNN and TNN workloads.
\end{itemize}

\section{Background}

\subsection{Binary and Ternary Neural Networks}

Traditional neural networks employ floating-point numbers for weights and activations.
A well-established method for reducing the energy cost of these workloads is operand quantisation,
i.e.,\ employing lower-bit representations such as 8-bit integers to trim down computational complexity and memory footprint.

\glspl{tnn} and \glspl{bnn} take this concept to its limit by representing both weights and activations with only three (-1, 0, 1) or two (-1, 1) values, respectively,
while maintaining acceptable accuracy levels in recognition and classification tasks~\cite{yuan_comprehensive_2023,li_ternary_weight_networks}.

In \glspl{bnn}, the quantisation process commonly uses the sign function \cite{yuan_comprehensive_2023}:
\begin{equation}
    q(x) = sign(x) = \begin{cases}
        +1, & \text{if } x >= 0,\\
        -1, & \text{otherwise}.
    \end{cases}
\end{equation}

For TNNs, a threshold function is used \cite{li_ternary_weight_networks}:
\begin{equation}
    q(x) = \begin{cases}
    +1, & \text{if } x > T, \\
    0,  & \text{if } |x| < T, \\
    -1, & \text{otherwise}.
    \end{cases}
\end{equation}
This work follows the methodology from \cite{li_ternary_weight_networks}, in which the threshold $T$ is calibrated independently for each layer.

\subsection{Analog Dot Products with Binary RRAM Crossbars}\label{sec:compute_modes}

In CNNs, the majority of computational effort is concentrated in the convolution layers\cite{lecun_deepl_2015}.
To implement these layers on a resistive CIM kernel, they are reformulated as matrix-vector multiplications (MVMs)~\cite{wei_lu_CIM_DNN_review_2024}.
The resulting MVMs are split into tiles and deployed to RRAM crossbars, where the weights are stored in memory cells.
A RRAM-based CIM-core of size $(N_\text{X} \times M_\text{X})$, as shown in Fig.~\ref{fig:crossbar}a, can process tiles of size $(M_t \times N_t)$, where $M_t <= M_\text{X}$ and $N_t <= N_\text{X}$.

This work focuses on binary \gls{rram} crossbars where the memristive cells are programmed to either \gls{lrs} or \gls{hrs}, and inputs are encoded as a train of binary pulses.
Those constraints have several advantages, such as better read margins, reduced ADC requirements and simpler programming circuits~\cite{shimeng_yu_bnn_rram_chip_2016}.
To deploy \gls{bnn} or \gls{tnn} workloads, weights and inputs in the set $\{-1,1\}$ or $\{-1,0,1\}$ must be encoded into the binary set $\{0,1\}$.

\subsubsection{Differential Weight Encoding}
In differential encoding, the weight values are represented with a pair of devices so that $w_n = \frac{1}{\Delta_I}(i^+_n - i^-_n$),
where $\Delta_I = I_{\text{LRS}} - I_{\text{HRS}} $ is the unitary step, a constant specific to the selected RRAM technology.
This is the preferred method for encoding weights, as it effectively represents zero-valued weights by cancelling out the non-zero HRS current~\cite{xiao_sandia_parasitics_2021,joshi_IBM_accurate_pcm_2020}.

As illustrated in Fig.~\ref{fig:crossbar}b, a pair of differential crossbar columns implements a dot product such that:
\begin{equation}
    \mathbf{x^b} \cdot \mathbf{w} = \sum_n^{N_\text{X}}{x^b_n w_n} = \frac{1}{\Delta_I} \sum_n^{N_\text{X}}{x^b_n\left(i^+_{n} - i^-_{n} \right)} = \frac{1}{\Delta_I}{\delta_{i}},
    \label{eq:dotproduct}
\end{equation}
where $\delta_{i} := f(\mathbf{x^b},\mathbf{w})$ is the current difference between the two columns,
 $\mathbf{x^b}$ is the encoded input vector in the binary set $\{0,1\}$, and $N_\text{X}$ is the column length.

\subsubsection{Input Encoding Alternatives}

Several input encoding methods can provide full accuracy, as the logic zero can be mapped to a pulse with \SI{0}{\volt} of amplitude.
Table \ref{tab:compute_modes} summarises commonly used alternatives\cite{parmar_bnn_compute_modes}.

Methods \emph{B-I} and \emph{B-II} use a shift-and-scale approach.
In \emph{B-I}, only positive inputs generate active voltage pulses, whereas in B-II, only negative inputs do.
Both methods require only a single dot-product cycle but apply only to BNNs.

\emph{T-I} employs differential encoding, representing each input value over two cycles: positive and negative.
\emph{T-II} uses bit-slicing based on the two's complement representation of the input and also requires two cycles.
Unlike \emph{B-I} and \emph{B-II}, both \emph{T-I} and \emph{T-II} can represent zero-value inputs, making them compatible with TNNs.

\begin{table}
    \centering
    \caption{Input Encoding Schemes}\label{tab:compute_modes}
    \begin{tabular}{|l|l|l|l|}
        \hline
        \bfseries Mode & \bfseries Input Mapping & \bfseries Dot-product $y = \mathbf{x} \cdot \mathbf{w}$ & \bfseries Cycles \\
        \hline \hline
        B-I   & $\mathbf{x} = 2\mathbf{x^+} - 1$  & $y = 2(\mathbf{x^+} \cdot w) - \sum_n^Nw_n$ & 1 \\
        \hline
        B-II  & $\mathbf{x} = -2\mathbf{x^-} + 1$  &$y = -2(\mathbf{x^-} \cdot w) + \sum_n^Nw_n$ & 1 \\
        \hline
        T-I   & $\mathbf{x} = \mathbf{x^+} - \mathbf{x^-}$ & $y = (\mathbf{x^+} \cdot w) - (\mathbf{x^-} \cdot w)$ & 2 \\
        \hline
        T-II  & $\mathbf{x} = -2\mathbf{x^1} + \mathbf{x^0}$ & $y = -2(\mathbf{x^1} \cdot w) + (\mathbf{x^0} \cdot w)$ & 2 \\
        \hline
    \end{tabular}
\end{table}

\section{Methods}

\subsection{Wire Parasitics}\label{sec:methods_parasitics}

Modeling the MVM output degradation caused by wire parasitics is paramount to assess the scalability of RRAM crossbars.
Previous work\cite{xiao_sandia_parasitics_2021} has stressed that the wire parasitics are expected to increase with technology down-scaling, which adds relevance to this issue.

A straightforward solution for estimating the crossbar dot-product currents under the presence of resistive wire parasitics is to solve a system of linear equations derived from nodal analysis.
However, for a crossbar of size $(N_\text{X} \times M_\text{X})$, this method has a complexity of $\mathcal{O}(M_\text{X}^3N_\text{X}^3)$~\cite{cao_parasitic-aware_2022}.
Hence, several heuristic approaches have been introduced.
The authors of~\cite{cao_parasitic-aware_2022} present a resistive crossbar model with very low error and a low time complexity of $\mathcal{O}(M_\text{X}N_\text{X})$, but it only works for passive crossbars, i.e.,\ without select transistors.

The open-source tool CrossSim~\cite{CrossSim} implements a heuristic method based on a successive under-relaxation algorithm.
Their implementation supports GPU acceleration, but it has an unbounded convergence time, and tuning the heuristic parameters is not straightforward.

This work sets its focus on the 1T1R topology shown in Fig.~\ref{fig:crossbar}, which has a high resilience against wire parasitics, as demonstrated by~\cite{xiao_sandia_parasitics_2021}.
We follow the modelling approach from~\cite{xiao_sandia_parasitics_2021} but implement a new circuit solver.
Instead of heuristics, we employ Algorithm~\ref{alg:compute_currents} to compute the output currents.
It is important to notice that the iteration loop (lines 5-7), which involves a chain of serial and parallel conductance reductions, is performed in parallel for all crossbar columns.
This vectorised algorithm yields a high throughput thanks to the strong use of Single-Instruction-Multiple-Data (SIMD) operations and multi-threading in Python's Numpy\cite{harris2020_numpy} library.
It has been verified against circuit simulations with SPICE and a maximum error of 0.15\% in the estimated output currents has been observed.

\begin{algorithm}[t]
    \caption{Compute Output Currents in a RRAM Crossbar}\label{alg:compute_currents}
    \begin{algorithmic}[1]
    \REQUIRE $G_\text{matrix} \in \mathbb{R}^{M \times N}$ (conductance matrix in $\mu$S), \\
             $input\_vector \in \{0,1\}^N$ (binary input vector), \\
             $R_\text{p}$ (parasitic resistance in $\Omega$), \\
             $V_\text{read}$ (read pulse amplitude in V)
    \ENSURE $I_\text{output} \in \mathbb{R}^M$ (output currents)
    \STATE \textbf{Mask} $G_\text{matrix}$: \\
           $active\_inputs \gets (input\_vector > 0)$ \\
           $G_\text{matrix\_gated} \gets G_\text{matrix}^\top$, \\
           $G_\text{matrix\_gated}[active\_inputs = 0, :] \gets 0$
    \STATE $g_\text{wire} \gets [{\frac{1}{R_\text{p}} \cdot 10^6}]_{M \times 1}$ \\
    \STATE $g_\text{per\_col} \gets \mathbf{0}_{M \times 1}$
    \FOR{$g_\text{row}$ \textbf{in} $G_\text{matrix\_gated}$}
        \STATE $g_\text{per\_col} \gets \frac{(g_\text{per\_col} + g_\text{row}) \cdot g_\text{wire}}{g_\text{per\_col} + g_\text{row} + g_\text{wire}}$
    \ENDFOR
    \STATE $I_\text{output} \gets g_\text{per\_col} \cdot V_\text{read}$
    \RETURN $I_\text{output}$
    \end{algorithmic}
\end{algorithm}

\subsection{ADC Calibration}\label{sec:methods_adc_calibration}

The output $d$ of an ideal ADC is given by
\begin{equation}
    d = \left\lfloor \frac{a}{\Delta_Q} + 0.5\right\rfloor,
    \label{eq:adc}
\end{equation}
where $a$ is the input signal, and $\Delta_Q$ is the quantisation step.

As explained in Section~\ref{sec:compute_modes}, interpreting the differential dot-product involves the unitary step $\Delta_I = I_{\text{LRS}} - I_{\text{HRS}} $.
Hence, achieving a full-precision reconstruction of this result requires a step size of $\Delta_Q = \Delta_I$ and a resolution of $B = \log_2{(N_\text{X}+1)}$ bits.

However, previous work has shown that \gls{cnn} workloads can operate without full precision~\cite{wei_lu_CIM_DNN_review_2024}.
Thus, the column dot-product result (\ref{eq:dotproduct}) can be approximated with:
\begin{equation}
    \mathbf{x} \cdot \mathbf{w} = \frac{1}{\Delta_I}{\delta_{i}} \approx s_l \left\lfloor \frac{\delta_{i}}{s_l\Delta_I} + 0.5\right\rfloor,
    \label{eq:mvm_out}
\end{equation}
where $s_l$ is a layer-wise scale parameter.

This work employs a calibration-based method to set the value of $s_l$ following this rule:
\begin{equation}
    s_l =
    \begin{cases}
    1, & \text{if } y^*_{l,\max} \leq 2^{b-1}-1 \\
    \frac{y^*_{l,\max}}{2^{b-1}-1}, & \text{otherwise},
    \end{cases}
    \label{eq:layer_scale}
\end{equation}
where $y^*_{\max}$ is the desired maximum value.

Increasing the ADC's step size extends its output range to meet the target range.
Simply put, this method trades off reduced clipping for a coarser quantisation error.
This is illustrated in Fig.~\ref{fig:crossbar}c.

To determine the target range, we simulate the inference using a small set of images from the training set.
The ADC outputs are monitored, and histograms are stored on disk.
Afterwards, we compute the mean $\mu_l$ and standard deviation $\sigma_l$ of the ADC outputs for each layer in the workload.
Finally we set $y^*_{\max}$ as:
\begin{equation}
    y^*_{\max} = \max{\big(|\mu_l - 3\sigma_l|, |\mu_l + 3\sigma_l|\big)}.
    \label{eq:layer_max}
\end{equation}

\subsection{CIM-Core Energy Model}\label{sec:methods_energy}

Previous work has shown that the energy consumed by MVM operations on RRAM crossbars is strongly dependent on the operand encoding,
matrix mapping and tiling schemes~\cite{cubero_rram_mvm_model_2024,andrulis_cimloop_2024}.

We implemented an additive, statistical energy model that uses the crossbar energy model from~\cite{cubero_rram_mvm_model_2024}.
Additionally, we consider the main mixed-signal peripheral circuits in the CIM-Core depicted in Fig.~\ref{fig:crossbar}a: Row Drivers (RD) and ADC.
Reference energy values for RD $E_{\text{RD}}$ are drawn from NeuroSim~\cite{peng_dnn_neurosim2_2021},
while for the ADC ($E_{\text{ADC}}$) these come from the analytical tool in \cite{andrulis_adc_2024}.
RRAM programming costs are not considered, as we are only interested in the run-time energy costs.

As explained in Section \ref{sec:compute_modes}, convolution and dense layers in CNN workloads are partitioned into tiles and mapped to crossbars.
We estimate the energy for each tile as:
\begin{equation}
    E_t = O_t \bigl(N_t \overline{\mathbf{x}}_t E_{\text{RD}} + M_t E_{\text{ADC}} + N_t M_t \overline{\mathbf{x}}_t \overline{\mathbf{g}}_t V_{\text{R}}^2 T_{\text{R}}\bigr),
    \label{eq:energy}
\end{equation}
where $O_t$ is the number of MVM operations done for tile $t$, $N_t$ is the number of active rows,
$M_t$ is the number of active columns,
$\overline{\mathbf{x}}_t \in [0,1]$ is the average input value,
and $\overline{\mathbf{g}}_t$ is the average cell conductance.
The last term in \ref{eq:energy} represents the energy spent in the memristive cells, which depends on the read voltage $V_{\text{R}}$ and effective pulse length $T_{\text{R}}$~\cite{cubero_rram_mvm_model_2024}.
The scaling factors $\overline{\mathbf{x}}_t$ and $\overline{\mathbf{g}}_t$ account for operand distribution statistics and the selected encoding scheme.

The average energy per MAC operation for a workload can be computed by dividing the tile energy by the number of MACs per tile ($O_t N_t M_t$) and summing over all tiles:
\begin{multline} \label{eq:energy_mac}
    E_{\text{MAC}} = \sum_{t}^{T} \bigg( \frac{E_t}{O_t N_t M_t} \bigg) = \\
    \sum_{t}^{T}\bigg( \frac{\overline{\mathbf{x}}_t E_{\text{RD}}}{M_t} + \frac{E_{\text{ADC}}}{N_t} + \overline{\mathbf{x}}_t \overline{\mathbf{g}}_t V_{\text{R}}^2 T_{\text{R}}\bigg).
\end{multline}

From this formula it is evident that to achieve a low energy per MAC, i.e.,\ a high energy efficiency, the number of active columns and rows per tile must be high.
Especially, a high number of active colums $N_t$ is important to amortise the elevated energy cost of ADC conversions.

\section{Results and Discussion}

\subsection{Exploration Framework}

To validate our work, we created an extended version of CIM-Explorer~\cite{pelke2025optimizingbinaryternaryneural},
an exploration framework consisting a TVM-based compiler and a CIM-core simulator.
The TVM-based compiler translates convolutional and dense layers to tiled MVM operations and offloads these to the CIM-core simulator~\cite{pelke2025optimizingbinaryternaryneural}.

The extended CIM-core simulator integrates all models presented in this work: the parasitic-aware current solver from Section~\ref{sec:methods_parasitics},
the ADC model with adjustable scale from Section~\ref{sec:methods_adc_calibration} and the energy model from Section~\ref{sec:methods_energy}.
It supports all operand mappings described in \ref{sec:compute_modes} and features workload statistics collection utilities.
All simulation components are implemented in Python and rely heavily on the Numpy\cite{harris2020_numpy} library to accelerate matrix operations.

\subsection{Workloads}

For our analysis, two neural networks were considered: a LeNet-5 network trained on the MNIST dataset and a VGG-7 network trained on Cifar10.
We used the open-source framework \emph{larq}~\cite{geiger_larq_2020} to train a binary and a ternary version of each network.
Both networks' architecture and training procedure are based on the implementation in \cite{li_ternary_weight_networks}.

Table~\ref{tab:workloads} summarises some relevant properties of these workloads.
It can be observed that the baseline top 1\% classification accuracy is higher for the \gls{tnn} implementations, thanks to their higher expressiveness~\cite{li_ternary_weight_networks}.
This difference is more pronounced for the more complex VGG-7 network than for the LeNet-5 network.

\begin{table}[]
\caption{Workloads}\label{tab:workloads}
\centering
\begin{tabular}{|l|l|l|l|l|}
    \hline
    \bfseries \multirow{2}{*}{Network} & \bfseries \multirow{2}{*}{Parameters} & \multicolumn{2}{c|}{\makecell{\bfseries Baseline Top 1\%\\ \bfseries Accuracy}} & \multirow{2}{*}{\makecell[tc]{\bfseries Average\\ \bfseries Matrix Size}} \\ 
    \cline{3-4}
                             &                             & \bfseries BNN                     & \bfseries TNN                     & \\
    \hline
    \hline
    LeNet-5                  & \num{93 322}                 & \SI{99.05}{\percent}              & \SI{99.35}{\percent}              & \num{63.6} $\times$ \num{286.4} \\
    \hline
    VGG-7                    & \num{18 286 986}             & \SI{90.18}{\percent}              & \SI{92.56}{\percent}              & \num{179.3} $\times$ \num{341.3} \\
    \hline
    \end{tabular}
\end{table}

\subsection{Operand Profiling}\label{sec:res_adc_calibration}

As the first step in our analysis, we create a profile of each workload.
For this purpose, we simulate the inference using 200 images and create frequency histograms of the workload operands and the ADC outputs.

\begin{figure}
    \begin{centering}
        \includegraphics[width=0.8\columnwidth]{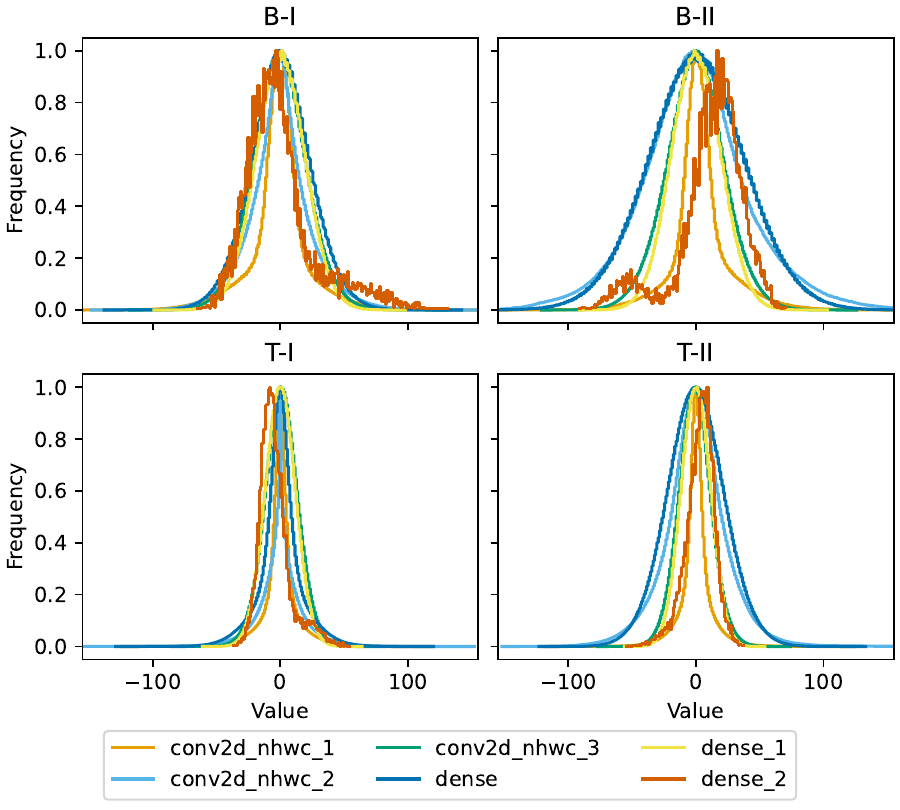}
        \caption{
            ADC output histograms for each layer of the VGG-7 network.
            Each subplot corresponds to a different input encoding scheme.
            Ternary mappings (T-I, T-II) yield narrower ADC ranges than binary mappings (B-I, B-II).
        }
        \label{fig:histograms_vgg7}
    \end{centering}
\end{figure}

As an illustration, the frequency histograms for each layer of the VGG-7 network are shown in Fig.~\ref{fig:histograms_vgg7}.
The differential weight encoding causes the ADC outputs to concentrate around zero.
Owing to the presence of zero-valued inputs in the TNN version of the workloads,
the ternary input encodings, \emph{T-I} and \emph{T-II}, yield a narrower ADC range than the binary encodings, \emph{B-I} and \emph{B-II}.

\subsection{Effects of Wire Parasitics}

\begin{table}
    \centering
    \caption{Considered Memristive Technologies}\label{tab:rram_techs}
    \begin{tabular}{|l|l|l|l|l|}
    \hline
    \bfseries Label & \bfseries Type & \bfseries LRS($\Omega$) & \bfseries HRS($\Omega$) & \bfseries Ref. \\  \hline \hline
    ReRAM-1    & ReRAM         & 1.00E+04 & 1.00E+05 & \cite{prezioso_training_2014}       \\ \hline
    PCM        & PCM           & 4.00E+04 & 1.76E+06 & \cite{joshi_IBM_accurate_pcm_2020}  \\ \hline
    ReRAM-2    & ReRAM         & 5.00E+04 & 4.00E+05 & \cite{yao_fully_hw_memr_cnn}        \\ \hline
    Perovskite & Charge-trapping  & 2.00E+05 & 2.50E+06 & \cite{chen_2024_perovskite}         \\ \hline
    IFG        & Floating-gate & 1.00E+07 & 2.00E+07 & \cite{fuller_ifg}                   \\ \hline
    \end{tabular}
\end{table}

\begin{figure}
    \begin{centering}
        \includegraphics[width=0.85\columnwidth]{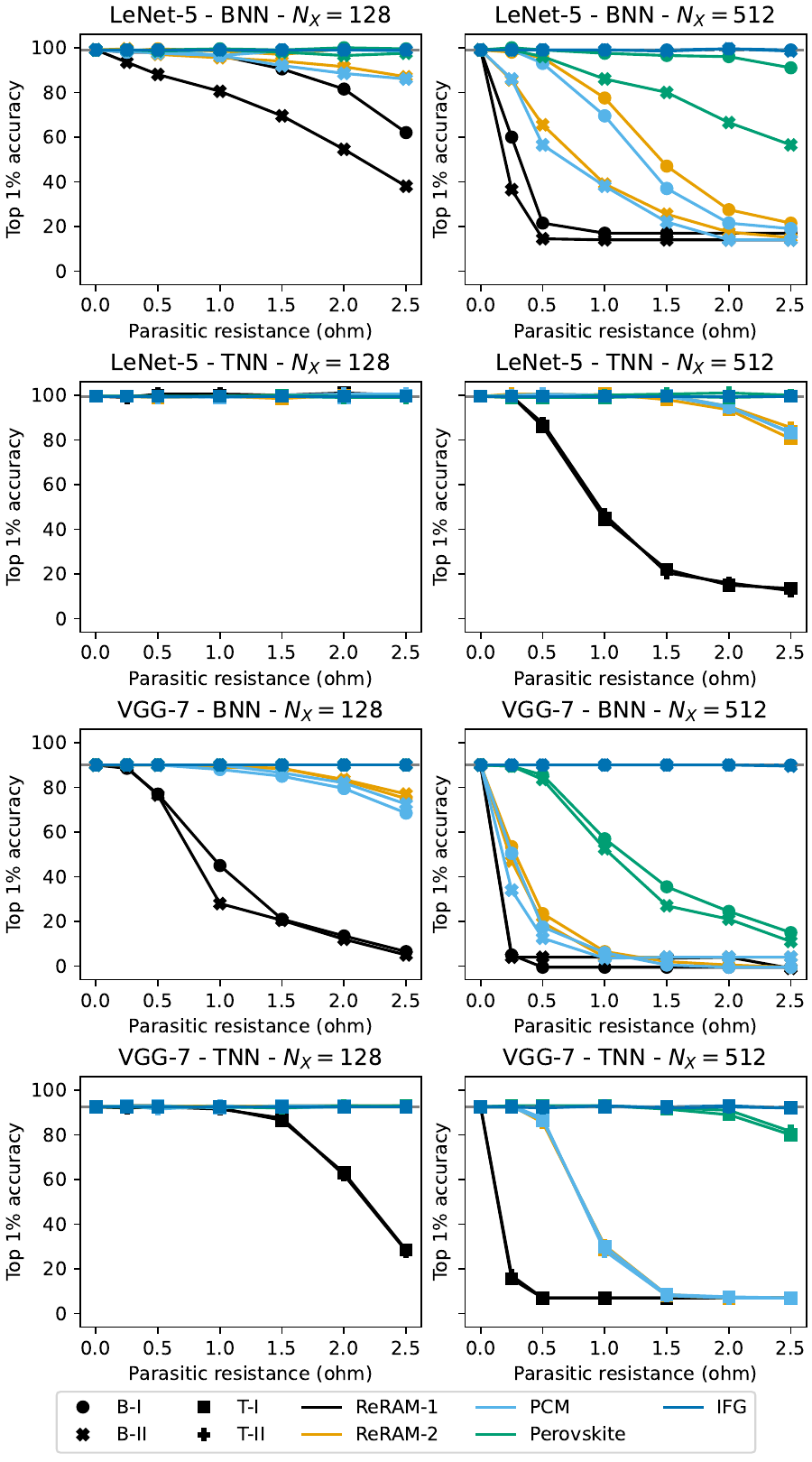}
        \caption{Impact of wire parasitics on accuracy.
        Each subplot presents a unique workload, quantisation and crossbar size combination.
        Markers indicate input encoding schemes, while colours distinguish memristive technologies.
        }
        \label{fig:res_parasitics}
    \end{centering}
\end{figure}

We ran a new set of inference simulations using another batch of 100 images to assess the resilience against crossbar wire parasitics.
We simulate with the \gls{lrs} and \gls{hrs} values of all technologies from Table~\ref{tab:rram_techs} and perform a sweep of the unitary wire resistance $Rp$ from \SIrange{0}{2.5}{\ohm}.
The obtained top 1\% classification accuracy is plotted in Fig.~\ref{fig:res_parasitics}.

The TNN workloads show a clear advantage against BNN, resulting from higher input sparsity.
In the case of LeNet-5-TNN, all memory technologies can maintain the baseline accuracy if the crossbar size is set to 128.
For the crossbar of size 512, two out of the five tested technologies can maintain the baseline accuracy.
VGG-7-TNN also performs better than VGG-7-BNN, but the observed degradation for this workload is more pronounced.

Due to their high LRS resistance, the device technologies labelled IFG and Perovskite show the best resilience against wire parasitics among all considered RRAM technologies.

\subsection{Evaluating the ADC Calibration Method}

We repeat the inference simulation using the same 100 images as in the previous section.
This time, the ADC resolution is varied from 8 to 3 bits, and the layer scale parameter $s_l$ is computed according to (\ref{eq:layer_scale}).
The wire parasitic resistance is set to zero to isolate the effect of ADC quantisation and clipping.

\begin{figure*}
    \begin{centering}
        \includegraphics[width=0.9\textwidth]{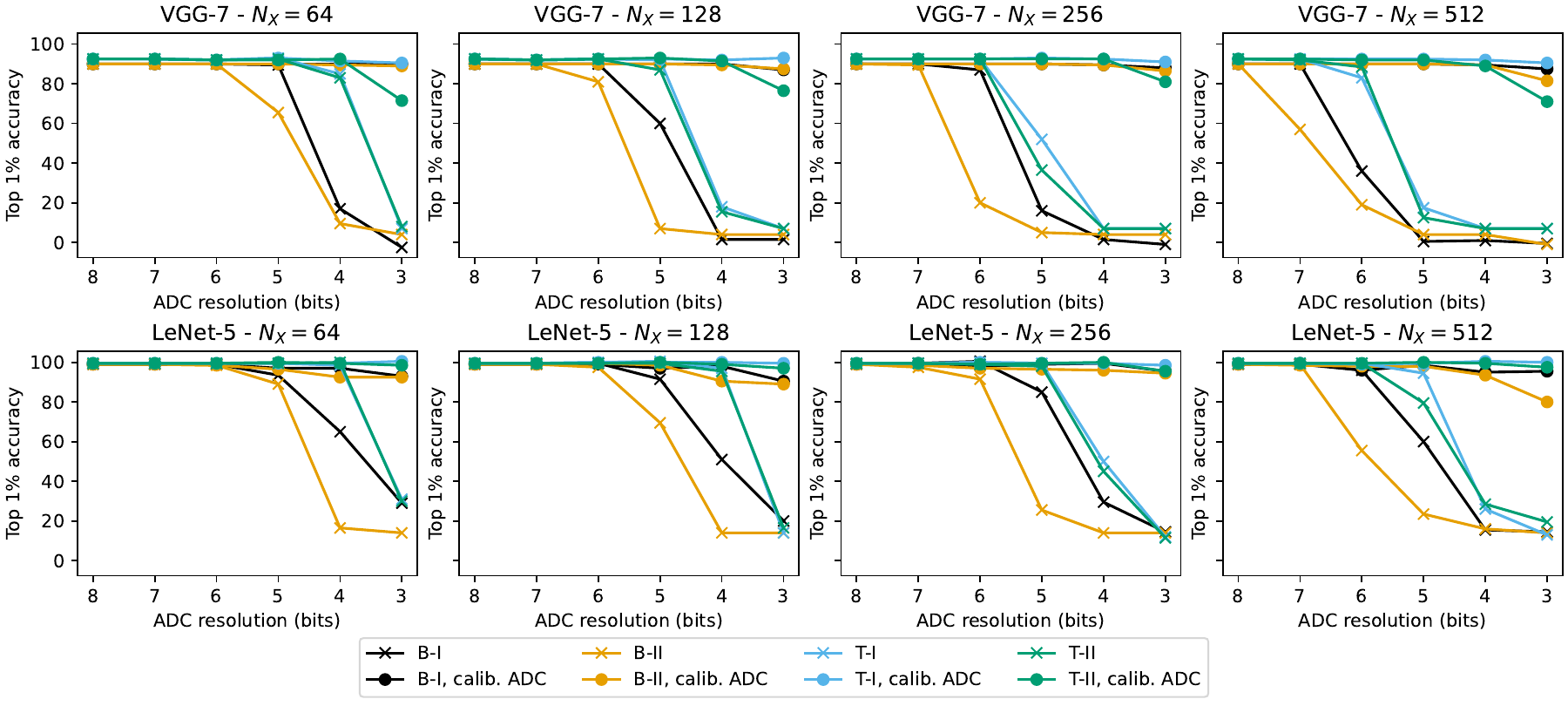}
        \caption{
         Accuracy vs. ADC resolution, comparing calibrated and uncalibrated settings.
         Calibration allows low-resolution ADCs with minimal accuracy loss.
         }
        \label{fig:res_adc_calibration}
    \end{centering}
\end{figure*}

The results of this experiment are shown in Fig.~\ref{fig:res_adc_calibration}.
The benefit of the ADC calibration method is clearly visible.
Without calibration, the LeNet-5 application requires 6-7 bits to maintain the baseline accuracy, and for VGG-7, 6-8 bits are required.
With calibration, 4 bits are enough for both applications and all considered input encodings.
The mapping \emph{T-I} shows the best performance among all input encodings.

\subsection{Energy Efficiency}

Finally, to round up our comparative analysis, we evaluate the energy efficiency of all considered workloads.
Mapping-dependent values ($N_t$ and $M_t$) and average input values ($\overline{\mathbf{x}_t}$) are computed from the operand histograms done in the profiling step (Section~\ref{sec:res_adc_calibration}).

\begin{figure}
    \begin{centering}
        \includegraphics[width=0.85\columnwidth]{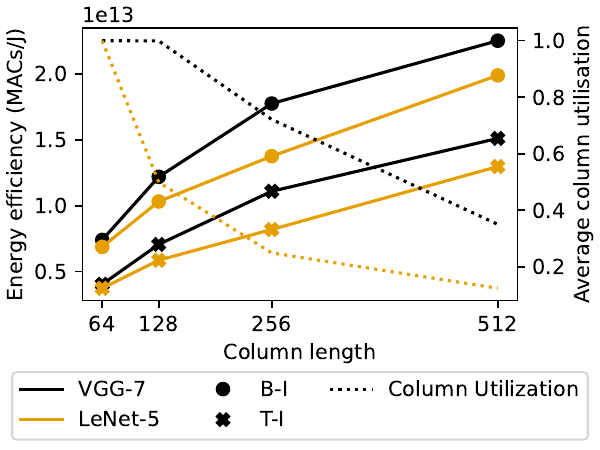}
        \caption{
            Energy efficiency and column utilisation vs. crossbar size.
        }
        \label{fig:res_energy}
    \end{centering}
\end{figure}

Our framework computes the energy efficiency in MAC/J for each workload and quantisation type using~(\ref{eq:energy_mac}).
The obtained energy efficiency for an ADC resolution of 4-bits and various crossbar sizes is plotted in Fig.~\ref{fig:res_energy}.
As expected, the energy efficiency increases together with the crossbar size.
However, the gains also diminish as the crossbar size grows.
This diminishing gain can be explained by looking at the decreasing column utilisation rate, also plotted in Fig.\ref{fig:res_energy}.
The difference in column utilisation also explains the superior energy efficiency of the larger VGG-7 network compared to the LeNet-5 network.

The high input sparsity in TNNs lowers the average input value $\overline{\mathbf{x}}_t$, reducing energy consumption per cycle.
However, the need for two dot-product cycles per operation decreases overall energy efficiency.
On average, the TNN versions of each workload show 40\% less energy efficiency than their BNN counterparts.

\section{Conclusion and Outlook}

We presented novel simulation models for evaluating the accuracy and energy efficiency of RRAM-based CIM accelerators,
focusing on crossbar scaling under the constraints of wire parasitics and limited ADC resolution.
Our parasitics model employs a fast vectorised algorithm with error rates below 0.15\%.
Using these models, we performed a quantitative comparison of BNN and TNN workloads.
Results show that due to their operand sparsity, TNNs require lower ADC resolutions and are less susceptible to degradation from wire parasitics.
For the studied networks, our ADC calibration method enables reducing ADC resolution to 4-bit with minimal accuracy loss.

Future work includes extending the energy model to incorporate other CIM-core components, such as input/output registers and post-processing arithmetic.
Additionally, alternative operand mappings and quantisation methods will be explored.

\bibliographystyle{IEEEtranS}
\bibliography{refs}

\end{document}